\documentclass[12pt]{article}

\def\D{{\cal{D}}}

\def\go{\;\&\;}

\def\ve{\vskip.5em}
\def\k{\kappa}

\def\vp{\4varphi}

\def\ora{\overrightarrow}

\def\Ra{\Rightarrow}

\def\half{\textstyle{\frac{1}{2}}}

\def\H{{\cal H}}

\def\vp{\varphi}

\def\H{{\cal H}}

\def\D{{\cal D}}

\def\ra{\rightarrow}
\def\tint{{\textstyle\int}}
\def\hg{{\hat g}}
\def\hp{{\hat\pi}}

\def\pa{\partial}
\def\dag{\dagger}

\def\b{\begin{eqnarray*}}  
\def\e{\end{eqnarray*}}    
\def\bn{\begin{eqnarray}}  
\def\en{\end{eqnarray}}   

\def\<{\langle}
\def\>{\rangle}

\def\no{\nonumber}

\def\ds{d^s\!x}
\def\k{\kappa}

\def\hk{\hat{\kappa}}
\def\{{\lbrace}
\def\hv{\hat{\varphi}}
\def\d3{d^3\!x}

\def\b{\beta}

\def\go{\;\&\;}

\def\Oz{{\cal{O}}}

\begin{document}

\title{   Quantum Physics has a \\
New, and  Remarkable, Expansion}

\author{John R. Klauder\footnote{klauder@ufl.edu} 
\\ 
University of Florida,   
Gainesville, FL 32611-8440}
\date{}
\let\frak\cal

\maketitle 

\begin{abstract} 
Canonical quantization has taught us great things.  A common example is that of the  harmonic oscillator, which is like swinging a ball on a string back and forth.  However, the half-harmonic oscillator blocks the ball at the bottom and then it quickly bounces backwards. This second model cannot be correctly solved using  canonical quantization.  Now, there is an expansion of quantization, called affine quantization, that can correctly solve the half-harmonic oscillator, and offers correct solutions to a grand collection of other problems,  which even reaches to field theory and gravity. 

This paper has been designed to introduce affine quantization; what it is, and what it can do.  
\end{abstract}

\section{Selected Features  of \\Quantum Mechanics}
There are rules that the classical variables, e.g., $p\go q$, need to follow when dealing with quantization. The most important requirement is that $-\infty<p\go q<\infty$. Certain functions of $p\go q$, are often accepted variables provided that they follow 
Poisson's bracket, which means  that  ${\bar{p}=f(p,q})$ while $\bar{q}=g(p,q)$, obeys 
  \bn \{\bar{q},\bar{p}\}\equiv (\pa\,g(p,q)/\pa q)(\pa f(p,q)/\pa p)-(\pa\,g(p,q)/\pa p)(\pa f(p,q)/\pa q)=1 \;.\en
  While we might find that $\{q,p\}=1=\{\bar{q},\bar{p}\}$, along with $-\infty<p\go q<\infty$,  may now find  $-\infty<\bar{p}\go \bar{q}<\infty$. They both have passed the requirements
   to create the quantum operators, $p\Ra P\go q\Ra Q$ along with   $\bar{p}\Ra \bar{P}\go \bar{q}\Ra \bar{Q}$, which they obey, $[Q,P]=i\hbar1\!\!1$ and $[\bar{Q},\bar{P}]=i\hbar1\!\!1$. While the bar-pair and the no bar-pair can be very different,  they were designed for the equality of $\bar{H}(\bar{p},\bar{q})=H(p,q)$, but find a {\it non}-equality in
  $\bar{\H}(\bar{P},\bar{Q})\neq \H(P,Q)$. The correct basic operators are the only ones to lead to valid results. How can we find the proper quantum operators? We now show how that is possible.
  
  We now introduce coherent states for canonical quantization which are defined by $
   |p,q\>=e^{-ipQ/\hbar}\,e^{iqP/\hbar}\!|\omega\>\equiv U(p,q)\,|\omega\>\,,$ 
  where we choose $[Q+iP/\omega]|\omega\> =0$, which leads to $\<\omega|P|\omega\>=\<\omega|Q|\omega\>=0$, while  $\<\omega|\omega\>=1$. It follows that $U(p,q)^\dag(aP+b\,Q)U(p,q)=a(P+p)+b(Q+q)$. We will find that coherent states provide a useful link between the classical and   quantum realms, specifically, a link between a classical Hamiltonian $H(p,q) $ and the correct quantum Hamiltonian $\H(P,Q)$.
  
  We next introduce the relation
  \bn  && H(p,q)=\<p,q|\H(P,Q)|p,q\>=\<\omega|\H(P+p,Q+q)|\omega\> \no \\ &&\hskip3.5em
  =\H(p,q)+\Oz(\hbar;p,q) \;.\en
  At this moment, the term $H(p,q)$ may have a hidden  dependence on $\hbar$, but that will disappear very soon.  Choosing the quantum Hamiltonian as a general polynomial, then the term $\H(p,q)$ is independent of $\hbar$. Therefore, letting  $\hbar\ra0$, it follows that $\Oz(\hbar;p,q)\ra0$, and $H(p,q)$ looses any $\hbar$ contribution.  Therefore, we find that now $H(p,q)=\H(p,q)\Ra \H(P,Q)$, showing how to pass from a classical function to the proper quantum  function. It is still possible that a special ordering of $\H(P,Q)$ is needed, but this is a standard issue. The final result is well known, but it has been derived because there will be a similar story below in which an
  operating order is a real issue, which  needs to be understood.
  
  The result in the last paragraph chose the classical variables, but how can that be the proper choice that leads to the correct Hamiltonian? Dirac has stated that the proper phase-space variables should be Cartesian, i.e.,  $d\sigma^2=\omega^{-1}\,dp^2+\omega\,dq^2$. The phase space does not include that relation, and a Cartesian requirement  is not to be found there. Indeed, this rule is semi-classical needing that $\hbar>0$. Using coherent states, a special metric, known as the  Fubini-Study metric,
  leads to
  \bn d\sigma^2=2\hbar\,[\,|\!|\,d|p,q\>|\!|^2-|\<p,q|\,d|p,q\>|^2\:]=\omega^{-1}\,dp^2+\omega\,dq^2\;,\en
  which has become Cartesian.  The minus sign term is designed to eliminate any phase factor, i.e. $|p,q;f\>=e^{if(p,q)}\,|p,q\>$, which is good physics. 
  
  Briefly stated, the coherent states have {\it created} the correct variables!

\section{A Standard Approach to \\Canonical Quantization}
The classical variables must be $-\infty<p\go q<\infty$ for  a  good reason. This permits the basic operators to be self-adjoint, i.e. $P^\dag =P\go Q^\dag=Q$, which is very important.

To understand what is self-adjoint, consider this simple example. For simplicity our example requires that
\bn \tint_a^b (d/dx) [f(x)\,g(x)]\;dx= f(b)g(b)-f(a)g(a)\;\;(\!=0)\;, \en
which we ask to be zero. This can be by choosing $g(a)=g(b)=0$, but that leaves $f(a)$ and $f(b)$ free to be anything. That is the case where $P^\dag\neq P$. However, if we need
 $P^\dag =P$, that requires $f(a)=f(b)= 0$ as well. For canonical quantization, the 
wave functions span the whole real line. In that case, where $P=-\hbar (\pa/\pa x)$, then 
\bn &&
\tint_{-\infty}^\infty \,P\,[\Phi(x)^*\Psi(x)]\;dx= -i\hbar\,[\Phi(x)^*\Psi(x)]\,|^\infty_{-\infty} =0\;\en
since Hilbert space requires finite elements , e.g., $\tint_{-\infty}^\infty\,|\Psi(x)|^2\;dx<\infty$, which forces each
wave function, $\Psi(x)$, to vanish at $x=\pm \infty$, i.e., $\Psi(-\infty)=\Psi(\infty)=0$.

What happens if the available  space  {\it does not span} 
the entire real line? Canonical quantization has adopted including a `virtual infinite 
wall', or just `v-wall'  for short, which is designed to squash wave functions to zero imitating, but not removing, and  simply ignoring that portion of space. 

As an example of that procedure, let us examine the harmonic oscillator with the classical Hamiltonian $H=(p^2+q^2)/2$. But now we require that $0<q<\infty$, for which it is now called the half-harmonic oscillator. Its classical behavior is that of a ball, hanging on a string, bouncing backwards at $q=0$. Now, using canonical quantization, the quantum Hamiltonian is still  $\H =(P^2+Q^2)/2 $, but the `v-wall' is located throughout $q<0$.  The region where $q>0$ is free to accept only the odd, not even, solutions of the harmonic oscillator that become zero at $q=0$, and then join  those eigenfunctions, with the portion that 
has been squashed,  to ensure a full, and
continuous function. After two derivatives it iwill again be zero at $q=0$, and thus there could be a second continuous function. That would be the accepted story in this case.

However, the first derivative, on the way toward the  second  derivative, is {\it not} a continuous function which immediately implies that the second derivative reaches infinity at the point $q=0$.
Such a wave function can not be part of any Hilbert space, because its normalization would be infinity. This problem will be correctly solved in the following section.

Another commonly studied example, using canonical quantization, is called `The particle in a box'. and its classical Hamiltonian is simply $H=p^2/2$.
In this example, the box consists of  $0<q<L<\infty$.
In this case, the customary procedure is to adopt {\it two} `v-walls', one throughout  $q<0$ and the other throughout $q>L$. By shifting the box to the space $-b<q<b<\infty$, for convenience, this example will  be considered in the next section.

\section{An Introduction to Affine Quantization}
The focus  is to seek examples without a full line coordinate  space,  such as the first example. 

\subsection{Choosing $0<q<\infty$}
In this example,   the point $q=0$ is removed, and  then discarding  $-\infty<q<0$,  while keeping $0<q<\infty$. Immediately, it follows that $P^\dag\neq P$.  That is accepted, and now introduce the dilation variable and also its coordinate, i.e.,  $d=pq\go q>0\Ra D=[P^\dag Q+Q P]/2=D^\dag\go Q=Q^\dag>0$. 
Hereafter, $D\go Q$ become the principal operators, not $P\go Q$, although $P$ will still have an important role to play.

The first usage of these variables is by examining the kinetic factor for classical Hamiltonians, i.e.,  $p^2=d^2/q^2$, in which 
\bn
d^2/q^2\Ra D(Q^{-2})D=P^2+(3/4)\hbar^2/Q^2 
=\hbar^2\,[\,-(d^2/dx^2)+(3/4)/x^2]\;.\en

We now introduce coherent states for affine  quantization which are defined by $
   |p;q\>=e^{ipQ/\hbar} e^{-i\ln(q)\,D/\hbar}\;|\beta\>\equiv U(p;q)\,|\beta\>\,,$ 
  where we choose $[(Q-1\!\!1)+i\,D/\beta\hbar]|\beta\> =0$, which leads to $\<\beta|Q|\beta\>=1$ and $\<\beta|D|\beta\>=0$, while  $\<\beta|\beta\>=1$. It follows that $U(p;q)^\dag(aD+b\,Q)U(p;q)=a(D+pqQ)+b(qQ)$. We will find that coherent states provide a useful link between the classical and   quantum realms, specifically, a link between a classical Hamiltonian-like  $H'(pq,q) $ and the correct quantum Hamiltonian-like  $\H'(D,Q)$.
  
  We now introduce the relation
  \bn  && H'(pq,q)=\<p;q)|\H'(D,Q)|p;q\>=\<\beta|\H'(D+pqQ,qQ)|\beta\> \no \\ &&\hskip4.1em
  =\H(pq,q)+\Oz'(\hbar;pq,q) \;.\en

The result in the last paragraph chose the classical variables, but how can that be the proper choice that leads to the correct Hamiltonian?  No 
  longer should they be  Cartesian, but, using the Fubini-Study metric, we  find that they are constant negative curvature with a vaue of $-2/\beta\hbar$, specifically it is $d\sigma'^2=(\beta\hbar)^{-1}\,q^2\;dp^2+(\beta\hbar)\;q^{-2}\,dq^2$. Once again, phase space does not include that relation, and now, our constant negative curvature requirement  is not to be found there. As was the prior case,  this rule is semi-classical, again needing that $\hbar>0$. 
  
Once again, and briefly stated, the affine  coherent states have {\it created} the correct variables! This treatment of canonical and affine quantization has been deliberately designed to be very   similar to prove they belong together.

These equations and their  variations are a fascinating feature of affine quantization. As a first example here, we adopt the classical Hamiltonian  $H=p^2/2+V(q)$, with $0<q<\infty$, it follows that 
the quantum Hamiltonian becomes
\bn \H(P,Q)=\half\,[\,P^2+(3/4)\hbar^2/Q^2]+V(Q)\;.
\en

As two important examples, we choose the harmonic oscillator and the half-harmonic oscillator. Adopting $m=\omega=1$ for simplicity, the classical harmonic oscillator Hamiltonian is $H_{ho}=(p^2+q^2)/2$, and using canonical quantization, the quantum Hamiltonian becomes $\H_{ho}=(P^2+Q^2)/2$. The eigenvalues of this operator are well known, and are $E_{n\:ho}=\hbar(n+1/2)$, for $n=0,1,2,...$. It is quite unique in having {\it equal spacing}, here at $1\hbar$.

Now, the half-harmonic oscillator has the identical classical Hamiltonian, $H_{h-ho}=(p^2+q^2)/2$, {\it 
but now}, $0<q<\infty$. This forces any particle to bounce off a `virtual-wall' at $q=0$ , and immediately turn around. The quantum Hamiltonian for the half-harmonic oscillator is given by 
$\H_{h-ho}=[P^2+(3/4)\hbar^2/Q^2+Q^2]/2$. The eigenvalues of this operator have been determined, and remarkably, they are $E_{n\;h-ho}=2\hbar(n+1)$, for $n=0,1,2,..$, and again they are {\it equally spaced}, now at $2\hbar$ 
\cite{a}. 
This usage of the number `2=twice' is everywhere, even in the fact that  
$-\infty<q<\infty$ is `twice' $0<q<\infty$! These remarkable results seem as if they both act like `pals'. 

It is well established that the harmonic oscillator has had a valid quantization. and now there is no reason to doubt that the half-harmonic oscillator also has a valid quantization.

\section{A Rich Realm of  
Examples Using \\Affine Quantization}
The harmonic operator story opened the door  to a great many more Hamiltonians and their promotions to valid operators.  For example,\ve
1. $0<q<\infty\go H=\half\,p^2+V(q)\Ra \H=\half[P^2+(3/4)\hbar^2/Q^2]+V(Q)$\ve
2. $-b<q<\infty \go H=\half\,p^2 +V(q)\Ra
 \H =\half[P^2+(3/4)\hbar/(Q+b)^2 +V(Q)$\ve
 3. $-b<q<b\go H=\half\,p^2\Ra \H=\half\,[P^2+\hbar^2(2Q^2+b^2)/(b^2-Q^2)^2]$\ve
 4. $0<b<|q|<\infty\go H=\half\,p^2+V(q)\Ra \H=\half\,[P^2+\hbar^2(2Q^2+b^2)/(b^2-Q^2)^2]+V(Q)$\ve
 5. $0<|q|<\infty\go H=\half\,p^2+V(q)\Ra \H= \half\,[P^2+2\hbar^2/Q^2]+V(Q)$\ve
 
 The term $P^2$ in each of these equations  is safe to say $P^2=-\hbar^2\,(d^2/dx^2)$ and $Q=x$, both in the standard Schr\"odinger representation. In  each one of these examples, the $q$-limitation becomes that of  an appropriate $x$-limitation.  The new $\hbar$-term in number 3. was derived in 
 \cite{b}.
 Note that number 5. comes from number 4. by letting $b\ra0$. It  means that only $q=0$ has been removed, but still $P^\dag\neq P $.
 
 Example 3. can be viewed as that of   `The particle in a box', which is often used as a simple teaching example, because it uses canonical quantization, and   $\cos$ and $\sin$ functions for its solutions. 
 As an example, the ground state for our box is $\cos(\pi\,x/2b)$ because at $x=\pm b$, it reaches zero to match its zero  wave function with that  of a
 `v-wall''. After 2 derivatives, leading again to a similar term, the  needed zero is there again. However, the first derivative, taken before the second derivative, encounters a non-continuous function that leads to an infinity, and which prevents that function  being accepted in any Hubert space. Briefly stated, canonical quantization fails `The particle in a box', while affine quantization succeeds. It may be helpful to define the eigenfunctions, which have the form  $\phi(x)=(b^2-x^2)^{3/2}\,f(x)$. Afterwards, focus on finding some solutions for $ f(x)$, which, at the present time (Nov. 2022), are still all unknown.
 
 \subsection { Introducing vector expressions}
 Just as a pointer, if there are vectors involved, it follows that in many places it is correct to   simply change single letters from $p\go q\ra \ora{p}\go \ora{q}$, and then   $P\go Q\ra \ora{P}\go \ora{Q}$.
 Now for $0<b^2<\ora{q}^2$ and $0<b^2<\ora{Q}^2$, we find that we have a  classical Hamiltonian, $H=\half\,\ora{p}^2+V(\ora{q})$, which leads to the quantum
  Hamiltonian,  $\H=\half[\ora{P}^2+\hbar^2\,(2\ora{Q}^2+b^2)/(\ora{Q}^
  2 -b^2)^2] +V(\ora{Q})$.

 \section{What Affine Quantization Can Do, and Has Already Done, for Physics}
 Applications of affine quantization are still developing, and this section will only  suggest
  reasons that affine quantization may be useful.
  A wider study of those proposals may be found in  several  of the author's articles, e.g.,  
  \cite{c,d,e}.

\subsection{Quantum field  theories, and  affine quantization} 
Conventional scalar field theories include classical Hamiltonians, such as
\bn H=\tint\{\half[\pi(x)^2+(\ora{\nabla}\vp(x))^2+m^2\vp(x)^2]+g\,\vp(x)^p\}\;\ds\;.\en
While the Hamiltonian $H<\infty$, this may still permit the fields to reach infinity. An example of that is $\pi(x)^2=A(x)/|x-x_0|^{s/2}$, with $A(x)>0$. Such behavior for fields of nature offers unwelcome  behavior, and such field infinites should be excluded. In that case, the affine variables can be helpful. \ve

{\bf Step 1} Following the expression $d=pq$ leads to the dilation field $\k(x)=\pi(x)\,\vp(x)$, which then 
leads to
\bn H=\tint\{\half[\k(x)^2/\vp(x)^2+(\ora{\nabla}\vp(x))^2+m^2\vp(x)^2]+g\,\vp(x)^p\}\;\ds\;.\en
Using these variables, $0<|\vp(x)|<\infty$ and $0\leq|\k(x)|<\infty$ so that $\pi(x) $ is well represented.                                    
As a bonus, it follows that $\vp(x)^p<\infty$ for every, even,  $p<\infty$, as well as $\ora{\nabla}{\k}(x)=(\ora{\nabla}{\pi}(x))(\vp(x
))+(\pi(x))(\ora{\nabla}{\vp}(x)))$, and the gradient term should exhibit no divergence lest it disturb the other terms.  {\it Nature is respected!}\ve

{\bf Step 2} Following the expression  $D=[P^\dag Q+Q P]/2 $  for $Q\neq 0$ leads to $\hk(x)=[\hp(x)^\dag \hv(x)+\hv(x)\,\hp(x)]/2$ and following $D(Q^{-2})D = P^2+ 2\hbar^2/Q^2$ leads to $\hk(x)(\hv(x)^{-2})\hk(x)=\hp(x)^2+2\hbar^2\,\delta(0)^{2s}/\hv(x)^2$. That  last term has many infinities because  several of Dirac's delta function,   i.e., $\delta(0)=\infty$,  designed so that  $\tint_{-a}^a \delta(x)\;dx=1$ for any $a>0$, 
and they need to be removed! \ve

{\bf Step 3}  Scaling is a tool of quantum field theory already, and treating $\delta(0) \equiv \D$,
temporarily as huge, but not infinity, we let scaling  work its magic. First, choose $\hp(x)\ra \D^{s/2}\;\hp(x)$, and then $\hv(x)\ra\D^{s/2}\;\vp(x)$. This leads us to  $\D\;^s\hp(x)^2+2\hbar^2\D^{2s}/\D^s\;\hv(x)^2$.   Finally,  multiply the  whole  last term by $\D^{-s}$, which will lead to
the first two factors in the  following quantum 
   Hamiltonian,  and which has added  the remaining  terms, using scaling, including that for  $g$,  as needed, and using Schr\"odinger's representation, to become
 \bn \H= \tint\{\half[\hp( x)^2 +2\hbar^2/\vp(x)^2+(\ora{\nabla}{\vp}(x))^2+m^2\,\vp(x)^2]+g\,\vp(x)^p\}\;\ds \:.
 \en 
 
 \subsubsection{Results from several Monte Carlo studies}
    At the start of affine Monte Carlo  (MC) investigations, around 2018, the factor   $2\hbar^2$ in the last equation,  
  was replaced by $(3/4)\hbar^2$, adopting a natural, but incorrect, classical reasoning. 
  However, the $(3/4)\hbar^2$ term  has performed very well.  The MC
results have been for  $\vp^p_n$ models, where $n=s+1$, and $s$ is the number of spatial dimensions, while  1 represents  time.  The model $\vp^4_4$, already examined in the 1980s, and started again around 2019, that  used canonical quantization, found that their  results were as if the interaction term was missing, while the similar studies, using affine quantization
found that their results were that the interaction term appeared as it should 
\cite{f}.
Additionally, the model $\vp^{12}_3$ had stronger positive results, as it should; see
\cite{g}.
Studies using $2\hbar^2$ should lead to even better results.

\subsection{Quantum gravity, and affine quantization}
This story follows the previous one in certain respects, which allow it to be shorter. The traditional classical variables are the momentum, $\pi^{ab}(x)$, and the metric, $g_{ab}(x)$. The dilation field  becomes $\pi^a_b(x)=\pi^{ac}(x)\,g_{bc}(x)$, with summation by $c$. All $a,b,c,
,...$ run over 1,2,3., and `$ab=ba$' in the fields. The metric field helps measure distance, such as $d\sigma^2(x)=g_{ab}(x)\;dx^a\;dx^b>0$ provided $\Sigma_{a=1}^3\:(dx^a)^2>0$. The quantum operators  then force $\hat{\pi}^{ab}(x)^\dag \neq\hat{\pi}^{ab}(x)$, because `$\hg_{ab}(x)>0$'. The dilation operator field is $\hp^a_b(x)=[\hp^{ac}(x)^\dag\hg_{bc}(x)+\hg_{bc}(x)\hp^{ac}(x)]/2$.
At this point, the classical ADM Hamiltonian 
\cite{n}, 
now in affine variables, is
\bn H=\tint\{ g(x)^{-1/2}[\pi^a_b(x)\pi^b_a(x)-\half\,\pi^a_a(x)\pi^b_b(x)]+ g(x)^{1/2}\,^{(3)}\!\!R(x)\}\;\ds\;,\en
where $g(x)=\det[g_{ab}(x)]$ and $^{(3)}\!\!R(x)$ is the Ricci scalar for 3 spacial coordinates.
The quantum ADM Hamiltonian , using Schr\"odinger's representation,  then becomes
\bn &&\H=\tint\{ [\hp^a_b(x) \,g( x)^{-1/2}\,\hp^b_a(x) -\half\,\hp^a_a(x) \,g( x)^{-1/2} \,\hp^b_b(x)] \no\\ &&\hskip12em +
g(x)^{1/2}\,^{(3)}\!\!R(x)\}\;\ds\;,\en
but there is a miracle in that $\hp^a_b(x)\;g(x)^{-1/2}=0$, for all  $a\go b$, and there are {\it no} Dirac delta functions as they appeared before.
There are constraints involved in a full quantization  of gravity,  for which there is special information  
\cite{j}.

\subsubsection{A comment regarding a path integration of gravity}

Since valid path integrations employ Wiener measures, an important element for a canonical field quantization would be $\omega^{-1}
\dot{\pi}(x,t)^2+\omega\,
\dot{\vp}(x.t)^2$, which ensures Cartesian behavior, while for
an affine field quantization it would be $\beta^{-1}
\vp(x,t)^2\,\dot{\pi}(x,t)^2+\beta\,\vp(x,t)^{-2}\,\dot{\vp}(x.t)^2$, which ensures that it  follows a constant negative  curvature.

Now, it happens  that the gravity Wiener measure element  needs to  be  $\gamma^{-1}\,[g_{ab}(x,t)\,\dot{\pi}^{ab}(x,t)]^2 +\gamma\,[g^{ab}(x,t)\,\dot{g}_{ab}
(x,t)]^2$, which also ensures a constant negative curvature, and an affine quantization, as   seen in
\cite{k}.

 \end{document}